\documentclass[aps,prd,onecolumn,showpacs,preprintnumbers]{revtex4}

\usepackage{graphicx}
\usepackage{amssymb}

\begin{document}


\title{Semiclassical quantization of gravity\\
       I: Entropy of horizons and the area spectrum}


  \author{T. Padmanabhan}\email{nabhan@iucaa.ernet.in}
  \affiliation{Inter-University Center for Astronomy and Astrophysics,
  Post Bag 4, Ganeshkhind, Pune-411007}
  \author{Apoorva Patel}\email{adpatel@cts.iisc.ernet.in}
  \affiliation{CTS and SERC, Indian Institute of Science, Bangalore-560012}


\date{\today}

\begin{abstract}
The principle of equivalence provides a description of gravity in terms
of the metric tensor and determines how gravity affects the light cone
structure of the space-time.
This, in turn, leads to the existence of observers (in any space-time)
who do not have access to regions of space-time bounded by horizons.
To take in to account this generic possibility, it is necessary to demand
that \emph{physical theories in a given coordinate system
must be formulated entirely in terms of variables
that an observer using that coordinate system can access}.
This principle is powerful enough to obtain the following results:
(a) The action principle of gravity must be of such a structure that,
in the semiclassical limit, the action of the unobserved degrees of freedom
reduces to a  boundary contribution $A_{\rm boundary}$ obtained by integrating 
a four divergence.
(b) When the boundary is a horizon, $A_{\rm boundary}$ essentially reduces
to a single, well-defined, term.
(c) This boundary term must have a quantized spectrum with uniform
spacing, $\Delta A_{boundary}=2\pi\hbar$, in the semiclassical limit.
Using this principle in conjunction with the usual action principle in
gravity, we show that:
(i) The area of any one-way membrane is quantized.
(ii) The information hidden by a one-way membrane leads to an entropy
which is always one-fourth of the area of the membrane, in the leading order.
(iii) In static space-times, the action for gravity can be given a purely
thermodynamic interpretation and the Einstein equations have a formal
similarity to laws of thermodynamics.    
\end{abstract}


\maketitle


\section{The Principle of Effective Theory}

Gravity is unique among all interactions in allowing a geometric description
of the space-time, with the components of the metric tensor $g_{ab}$ as the
fundamental variables which describe the gravitational interaction.
Given the torsion-free Christoffel symbols $\Gamma^a_{bc}$, obtained from
the metric tensor, it is possible to to define a local inertial frame of
``first order'' accuracy around any event ${\cal P}$, in which the metric
tensor reduces to its Minkowski form and the first derivatives of the metric
tensor vanish.
We shall take the principle of equivalence to imply that
\emph{the laws of special relativity are valid in this local inertial frame}.
This allows one to determine the interactions of gravity to other fields by
expressing them in a generally covariant manner (using the``comma-to-semicolon
rule'', say) in the local inertial frame, and then extending them to curved
space-time.
    
By studying the propagation of null geodesics in a space-time, one can define
null surfaces (``light sheets") which cannot be crossed by any material
particle or signal. 
It follows that, in a space-time with a non-trivial metric tensor $g_{ab}(x^k)$,
there could exist a light cone structure such that information from one region
is not accessible to observers in another region. 
It should be stressed that
such a limitation  is \emph{always}  observer/coordinate dependent.
(Throughout this paper, we use the term ``observers" in the sense of
a well-defined family of time-like curves.)
To appreciate this fact, let us begin by noting that the freedom of choice
of the coordinates allows 4 out of 10 components of the metric tensor to be
pre-specified.
We take these to be $g_{00}=-N^2,~g_{0\alpha}=N_\alpha$,
though this choice is by no means unique.
(We use the space-time signature $(-,+,+,+)$ and other sign conventions of
ref.\cite{mtw}. Whenever not explicitly mentioned, our units are $G=\hbar=c=1$.
The Latin indices vary over 0-3, while the Greek indices cover 1-3.)
These four variables characterize the observer dependent information.
For example, with the choice
$N=1,N_\alpha=0,g_{\alpha\beta}=\delta_{\alpha\beta}$, the ${\bf x}=$
constant lines represent a class of inertial observers in flat space-time,
while with $N=(ax)^2,N_\alpha=0,g_{\alpha\beta}=\delta_{\alpha\beta}$,
the ${\bf x}=$ constant lines represent a class of accelerated observers
with a horizon at $x=0$.
\emph{We  only need to change the form of $N$ to make this transition,
whereby a class of time-like trajectories, ${\bf x}=$ constant,
acquire a horizon.} 
Similarly, observers plunging in to a Schwarzschild black hole will find it
natural to describe the metric in the synchronous gauge, $N=1,N_\alpha=0$
(see e.g., ref.\cite{ll}, \S97),
in which they can access the information present inside the horizon.
On the other hand, the more conservative observers follow the $r=$constant
($>2M$) lines in the standard foliation, which has $N^2=(1-2M/r)$,
and the surface $N=0$ acts as the horizon which restricts the flow of
information from $r<2M$ to the observers at $r>2M$.
This approach treats the coordinates $x^a$ as markers
and $g_{ab}(x)$ as field variables; the gauge transformations
of the theory allow changing the functional forms of $g_{ab}(x)$,
in particular those of $g_{00}(x)$ and $g_{0\alpha}(x)$.
The ${\bf x}$=constant trajectories provide the link between
the gauge choice and the conventional concept of a class of observers.
In any given gauge, there could exist a class of time-like trajectories that
are not necessarily geodesics (i.e. observers who are not freely falling),
which have only restricted access to regions of space-time.

This aspect, viz. that different observers may have access to different
regions of space-time and hence differing amount of information,
introduces an unusual feature in the theory.
There is a strong motivation to postulate that
\emph{physical theories in a given coordinate system must be formulated
entirely in terms of the variables that an observer using that coordinate
system can access}.
We call this postulate the ``principle of effective theory''.
The detailed implications of this postulate will be elaborated up on as we
go along.

This postulate is a new addition to the traditional principles of general
relativity, although it is a familiar principle in high energy physics.
In the simplest context of field theories, a principle of the above kind
``protects" the low energy theories from the unknown complications of the
high energy sector. For example, one can use QED to predict results
at, say, $10$ GeV without worrying about the structure of the theory at
$10^{19}$ GeV, as long as one uses coupling constants and variables
defined around $10$ GeV and determined observationally.
In this case, one invokes the effective field theory approach in the
momentum space, and the effect of high-energy modes is essentially
absorbed in the definitions of field variables and coupling constants
appearing in the low energy Lagrangian.
The powerful formalism of renormalization group describes how a theory
changes as its domain of applicability is changed, and the symmetries of the
theory often provide a good guide to the types of changes that may occur.
We have learned from experience that in effective field theories
everything outside the domain accessible to the observer is reduced to:
(a) change of variables, (b) change of couplings,
(c) higher derivative interaction terms and (d) boundary terms.
(The number of interaction terms in the effective field theory may become
infinite, but they can be organized in increasing powers of derivatives).
Our postulate invokes the same reasoning in coordinate space (which is
commonplace in condensed matter physics and lattice field theories),
and demands---for example---that the observed physics outside a black hole
horizon must not depend on the unobservable processes beyond the horizon.
The effective field theory of gravity must be obtained by summing over
different configurations---and possibly different topologies---beyond the
horizon.
In such a theory, among the possible changes listed above,
(a) is unimportant because the geometrical description of gravity
identifies the metric tensor as the natural fundamental variable;
(b) is unimportant because the couplings are fixed using the effective theory;
and (c) is unimportant at the lowest order of effective theory.
Hence we concentrate on the boundary terms in the theory, i.e.
we expect the action functional describing gravity to contain certain
boundary terms which are capable of encoding the information equivalent
to that present beyond the horizon. 

While we introduce the principle of effective theory as a postulate,
we would like to stress that it is a fairly natural demand.
To see this, let us recall that in the standard description of flat
space-time physics, one often divides the space-time by a space-like
surface $t=t_0$=constant.
With appropriate information on this surface, one can predict the
evolution for $t>t_0$ \emph{without knowing the details at} $t<t_0$.
In case of curved space-times with horizons, similar considerations apply.
For example, if the space-time contains a Schwarzschild black hole,
then the light cone structure guarantees that the processes inside the
black hole horizon cannot affect the outside events \emph{classically},
and our principle of effective theory is trivially realized.
But the situation in \emph{quantum theory} is more complicated because
quantum fields can have non-trivial correlations across the horizon
and---in general---can lead to processes which are classically forbidden.
For example, the ground state wave functional of a scalar field at $t=0$
in flat inertial coordinates $(t,{\bf x})$ can be expressed as
$\langle 0| \phi_L({\bf x}),\phi_R({\bf x}) \rangle$,
where $|0\rangle$ is the inertial vacuum state and 
$| \phi_L({\bf x}),\phi_R({\bf x}) \rangle$ is a quantum state with the field configuration
$\phi_L({\bf x})$   at $x<0$ and the field configuration
 $\phi_R({\bf x})$  at $x>0$.
Observers who have no access to $x<0$ region will describe the same state
with a density matrix obtained by integrating out the $\phi_L({\bf x})$.
If this lack of accessibility arises due to using a gauge
in which $N=0$ on the intervening boundary,
then the resulting density matrix will be thermal \cite{lee},
with a temperature determined by the first derivative of $N$ on the boundary.
This effect arises essentially due to the nontrivial quantum correlations
which exist across the boundary in the ground state
$\langle 0| \phi_L({\bf x}),\phi_R({\bf x}) \rangle$.
Even in the absence of matter fields, correlations of quantum fluctuations
of gravity (treated as spin-2 modes propagating in the classical metric)
can lead to entanglement of modes across the horizon.
 
The principle of effective theory implies that it must be possible to
\emph{protect} the physical processes outside the horizon from 
such quantum effects across the horizon.
Since the horizon is the only common element to inside and outside regions,
the effect of these entanglements across a horizon can only appear as a
boundary term in the action.
Moreover, if this relic of quantum entanglement survives in the classical
limit, it must be expressible as a total divergence so as not to affect the
classical equations of motion.
Hence it is an inevitable consequence of principle of effective theory that
\emph{the quantum action functional describing gravity must contain certain
boundary terms}, arising out of integrating total divergences over the bulk,
which are capable of encoding the information equivalent to that present
beyond the horizon. 

This framework imposes a very strong constraint on the form of action
functional $A_{grav}$ that describes semiclassical gravity.
To study the effects of unobserved degrees of freedom in some space-time
region, let us  divide the space-time manifold in to two regions separated
by a boundary surface.
We choose a coordinate system such that this boundary acts as a horizon for
the observer on one side (say side 1), which usually requires non-trivial
values for the gauge variables $N,~N_\alpha$. (For example, many space-times
with horizons---Schwarzschild, de Sitter etc.---admit a coordinate chart with
$N^2=g^{rr}$ with $N^2$ having a simple zero at the location of the horizon.)
In quantization of gravity, attempted  through the functional integral
approach based on the Lagrangian, a sum over all paths with fixed boundary
values provides the transition amplitude between initial and final states.
This path integral approach is more suitable than the Hamiltonian approach
(based on non-commuting operators) for dealing with effective theories,
since it uses the action functional directly and allows easy integration
over unobserved degrees of freedom.
The effective theory for the observer on side 1 is obtained by
integrating out the variables on the inaccessible side (side 2).
With a local Lagrangian, it is formally given by
\begin{equation}
\exp [iA_{\rm eff}(g_1)] \equiv \int [{\cal D} g_2] ~
\exp [i(A_{\rm grav}(g_1) + A_{\rm grav}(g_2))] ~.
\end{equation} 
In this integration, the intrinsic geometry of the common boundary has to
remain fixed, e.g. by choosing $g$ to remain continuous across the boundary.
In the semiclassical limit, the integration over $g_2$ can be done by
saddle-point approximation.
The result is a product of exponential of the classical action
and the determinant of small quantum fluctuations.
The effective theory on side 1 is thus described by the
action $A_{\rm eff}^{\rm WKB}(g_1)$, with 
\begin{equation}
\exp [iA_{\rm eff}^{\rm WKB}(g_1)] =
\exp [i(A_{\rm grav}(g_1) + A_{\rm grav}(g_2^{\rm class}))]
\times {\rm det}(Q) ~.
\end{equation}
This result has several non-trivial implications:

(i) Since we expect the effects of unobserved degrees of freedom to be
described by a boundary term, we get the constraint that, when evaluated
on the classical solution, the action $A_{\rm grav}(g_2^{\rm class})$
must be expressible in terms of the boundary geometry.
That is, $A_{\rm grav}(g_2^{\rm class})=A_{\rm boundary}(g_1)$, and
\begin{equation}
\exp [iA_{\rm eff}^{\rm WKB}(g_1)] =
\exp [i(A_{\rm grav}(g_1) + A_{\rm boundary}(g_1))]
\times {\rm det}(Q) ~.
\end{equation}
This is a  nontrivial requirement.
For example, the standard Lagrangian for a scalar field,
$L=-\frac{1}{2}\partial_a \phi\partial ^a\phi - V(\phi)$,
does not lead to an action which satisfies this criterion.
When evaluated on the classical solution,
$-\partial_a \partial^a \phi + V'(\phi) = 0$,
the action becomes
\begin{equation}
A_{\rm scalar}^{\rm class} =
\int d^4x \left( \frac{1}{2} \phi V'(\phi)-V\right)
- \frac{1}{2} \int d^4x~\partial_a \left( \phi \partial^a \phi \right) ~.
\label{scalar}
\end{equation}
The second term is a total divergence, but not the first
(except in the trivial case of a free scalar field with $V(\phi) \propto \phi^2$).

(ii) Since the boundary term arises  due to the choice of a specific
coordinate system, in which the boundary acts as a one-way membrane,
$A_{\rm boundary}(g_1)$ will in general depend on the gauge variables
$N,N_\alpha$ and will \emph{not} be generally covariant.
Classically, with the boundary variables held fixed, the equations of
motion remain unaffected by a total divergence boundary term;
the fact that the boundary term is not generally covariant is
unimportant for classical equations of motion.
But even a total divergence boundary term can affect the quantum dynamics,
because quantum fluctuations around classical solutions can sense the
properties of the boundary.
We emphasize that the quantum theory is governed by $\exp[iA_{\rm eff}]$
and not by $A_{\rm eff}$.
The boundary term, therefore, will have no effect in the quantum theory,
if the quantum processes keep $\exp[iA_{\rm boundary}]$ single-valued.
This is equivalent to demanding that the boundary term has a discrete
spectrum, with uniform spacing $\Delta A_{\rm boundary}=2\pi$.

(iii) Since the boundary term arises because of our integrating out the
unobserved degrees of freedom, the boundary term should represent the loss
of information as regards the particular observer (encoded by the choice
of gauge variables $N,~N_\alpha$) and must contribute to the entropy.
This strongly suggests that, in the Euclidean sector, the effective action
$A_{\rm eff}^{\rm WKB}(g_1)$ must have a thermodynamic interpretation.
Indeed, there is a deep connection between the standard thermodynamic
results obtained for quantum fields in curved space-time and the nature
of semiclassical gravity.

\section{Role of Gauge Transformations in Gravity}

Before we start the detailed analysis,
it is important to clarify some strategic issues in our approach.
Our description treats the gravitational degrees of freedom and the concept
of general covariance in a manner different from the conventional approaches,
and we stress the following: 

(a) The geometrical interpretation of gravity treats $g_{ab}$ as a metric
tensor living on a manifold, and its specific form (in ``index notation")
depends on the coordinates chosen on the relevant region of the manifold.
By using intrinsically geometrical constructs, classical gravity ensures
that the choice of coordinates is unimportant. In that approach,
the changes in $g_{ab}$ are specifically related to changes of coordinates.
The field theoretic description of gravity, on the other hand,
is conceptually different. Here, one chooses the coordinate system to be
specified a priori and allows for certain well defined gauge transformations
of the the fundamental variables $g_{ab}$ to be invariances of the theory.
The transformation that changes $g_{00}=-1$ to $g_{00}\ne -1$,
for example, could arise as such a gauge transformation.
The field theoretical and geometrical view points
lead to same observable results in classical theory.
In particular, if a class of time-like trajectories have restricted access to
regions of space-time, that result will continue to hold in both descriptions.
In the geometrical language, one will introduce a coordinate transformation
which is natural to the class of observers (say by Fermi-Walker transport),
and the components of the metric will change tensorially under this
transformation.
In the field theoretic language, one changes the gauge used for the
description, thereby changing the form of metric components.
In quantum theory, the second approach seems more natural, especially
since quantum theory has certain level of observer dependence built in to it.
To each class of observers, we can associate a natural gauge
(somewhat like the non-covariant transverse gauge in electrodynamics,
$\phi=0, \partial_\alpha A^\alpha=0$, which needs to be changed for
every Lorentz observer), and we would like to have a totally
self-contained description of physics for each observer. 
Our fiducial observers can always be taken to follow the timelike
trajectories ${\bf x}=$ constant in a region of space-time, and depending
on the gauge choice they may have access to part or whole of space-time.
It is possible that certain manifolds allow some kind of ``global observers",
and an associated gauge which is well defined throughout the manifold
(e.g. the inertial observer in flat space-time or the Kruskal observers
in Schwarzschild space-time).
But this is a luxury we shall not rely up on,
because the principle of effective theory demands complete description
of physics in different gauges without relying on a global description.

(b) The existence of a one-way membrane is \emph{always} an observer dependent
statement, where we interpret ``observer" as a family of time-like world-lines.
There exist observers in black hole space-time who \emph{can} access the
information inside the horizon, and there are observers in flat space-time
who do not have access to all the information.
As explained earlier, the form of the 4 gauge variables
allows us to characterize the observer dependent situations.
The division of space-time by a horizon uses a specific gauge,
and the resultant $A_{\rm boundary}(g_1)$ in general depends
on the gauge variables $N,N_\alpha$.
The precise definition of a ``horizon" or a ``one-way membrane" is
mathematically intricate, but we do not need it.
There is fair amount of literature (see eg. \cite{date})
on the definitions of different classes of horizons
(event, apparent, Killing, dynamic, isolated, etc.),
but fairly simple definitions are adequate for our purpose.
For example, we shall often use the criterion that a boundary defined by
$N=0$, with all other metric components well behaved near it,
is a one-way membrane.
The fact that such a criterion is not ``intrinsic" or ``geometrical"
is irrelevant, since we do not make any distinction of principle between,
say, Rindler horizon and Schwarzschild horizon.

(c) Of the gauge variables $N,N_\alpha$, the lapse function $N$
plays a more important role in our discussion than $N_\alpha$,
and we can set $N_\alpha=0$ without loss of generality. 
To explicitly see how $N$ can change,
consider the \emph{infinitesimal} space-time transformation
$x^i \to x^i + \xi^i(x^j)$, with the condition
$g_{\alpha\beta}\dot\xi^\beta= N^2 (\partial \xi^0/\partial x^\alpha)$,
which is equivalent to
\begin{equation}
\xi^\alpha = \int dt~N^2 g^{\alpha\beta}
              \frac{\partial\xi^0}{\partial x^\beta} + f^\alpha(x^\beta) ~.
\label{largegauge}
\end{equation}
Such transformations keep $N_\alpha=0$, but change $N$ and $g_{\alpha\beta}$
according to $\delta g_{ij} = - \nabla_i\xi_j - \nabla_j\xi_i$ (see  ref.\cite{ll}, \S97).
With the extra conditions $N=1, \dot\xi^0=0$,
the class of transformations specified by the four functions $\xi^i(x^j)$
change one synchronous reference frame to another.
(As an aside, we mention that there is a deeper dynamical reason
as to why $N$ plays a more important role than $N_\alpha$.
In the Hamiltonian formulation of general relativity,
$N$ and $N_\alpha$ play the role of Lagrange multipliers in the action
functional, and their variation leads to the $00$ and $0\alpha$ components
of the Einstein equations.
It is, however, possible to set $N_\alpha=0$,
work out the algebra of primary constraints,
and recover the $0\alpha$ equation as a consistency condition for closure.
In this sense, $N_\alpha$ is  less important than $N$ even dynamically.
Nevertheless, our results are independent of this condition.)

(d) The {\it infinitesimal} gauge transformations of the kind
$\delta g_{ij} = - \nabla_i\xi_j - \nabla_j\xi_i$,
induced by the four gauge functions $\xi^i$, are close to identity
and the results quoted above are valid to first order in $\xi$.
For example, the transformation induced by $\xi^a_{(R)}=(-aXT,-(1/2)aT^2,0,0)$
changes the flat space-time metric $g_{ab}=(-1,1,1,1)$ to the form
$g_{ab}=(-(1+2aX),1,1,1)$, up to first order in $\xi$.
Obviously, one cannot consider a situation in which $N\to 0$
within the class of infinitesimal transformations.
The theory, however, is also invariant under finite transformations,
which as we shall see are more ``dangerous".
Of particular importance are the ``large gauge transformations",
which are capable of changing $N=1$ in the synchronous frame
to a nontrivial function $N(x^a)$ that vanishes along a hypersurface.
Well-known examples are the transformation from the synchronous frame for the
Schwarzschild space-time to the standard coordinate system with $N^2=(1-2M/r)$,
and the transformation from the inertial coordinates in flat space-time to
Rindler coordinates with $N^2=(ax)^2$.
In all these cases, the $N=0$ surface is the horizon for the ${\bf x} = $ constant
 observers in that frame.
In particular, the coordinate transformation $(T,{\bf X})\to(t,{\bf x})$ in the region $|T| < X$ of the 
Minkowski spacetime: 
\begin{equation}
1+aX=(1+ax)\cosh at; \quad aT=(1+ax)\sinh at;\quad Y=y;\quad Z=z
\label{stdrindler}
\end{equation}
changes the metric from $g_{ab}=(-1,1,1,1)$ to $g_{ab}=(-(1+aX)^2,1,1,1)$.
(The infinitesimal version of this transformation is generated by $\xi^a_{(R)}$
in the limit of small $aX$.)
Given such large gauge transformations,
we can discuss regions arbitrarily close to the $N=0$ surface,
which is the Rindler horizon at $x_H=-1/a$.
The importance of this transformation lies in the fact that a very wide
class of horizons can be approximated by the metric in this gauge
$g_{ab}=(-(1+aX)^2,1,1,1)$ close to the horizon.
If the metric is described in a gauge in which $g_{\alpha\beta}$ are finite,
and well behaved near the hypersurface $x=x_H$, then by expanding $N$ in a
Taylor series as $N=N'(x_H)(x-x_H)+\ldots$ and diagonalising $g_{\alpha\beta}$,
we obtain the metric close to the horizon in the Rindler form.
In fact, the transformation in (\ref{stdrindler}) generalizes
in a simple manner to cover even the situation
where the $N=0$ hypersurface is time dependent.
The transformation in this case is given by
(see e.g. section 6 of \cite{tpwork1})
\begin{equation}
X = \int^{\prime} \sinh \mu (t) dt + x \cosh \mu (t); \qquad 
T = \int^{\prime} \cosh \mu (t) dt + x \sinh \mu (t)
\label{eqn:fiftyfive}
\end{equation}
The form of the metric in this gauge is remarkably simple,
\begin{equation}
ds^2 = - (1+g(t)x)^2 dt^2 + dx^2 + dy^2 + dz^2 ~,
\label{eqn:fiftyseven}
\end{equation}
where the function $\mu(t)$ is related
to the time dependent acceleration $g(t)$, by $g(t) = (d\mu/dt)$.
This transformation is a good approximation close to any time dependent
horizon with $x_H(t)=-1/g(t)$.
In particular, many particle horizons which arise in cosmological models
can be approximated---close to the horizon---by such metrics.

(e) These gauges, which are singular on the horizon will play an important
role in our discussion. Their significance arises from two principle reasons.
When analytical continuation to Euclidean time is carried out,
the horizon hypersurface reduces to a singular hypersurface,
pinching the space-time and opening up possibilities of interpretation in terms of
topology change
\cite{chrisduff,tpinstanton,inprogress}.
The transformations that make $N$ vanish, therefore, have serious
topological implications, and can change the value of $A_{\rm boundary}(g_1)$.
More directly, the symmetries of the theory enhance significantly near the
$N=0$ hypersurface.
An interacting scalar field theory, for example, reduces to a $(1+1)$
dimensional CFT near the horizon, and the modes of the scalar field vary as
$\phi_{\omega} \cong \vert x - x_H\vert^{\pm (i\omega / 2a)}$
with $N'(x_H)=a$ (see e.g. section 3 of \cite{tpwork1}).
Similar conformal invariance occurs for gravitational sector as well.
If the near horizon metric is written in the gauge, 
\begin{equation}
ds^2 = -ax \ dt^2 + \frac{dx^2}{ax} + \sigma_{AB}(x,x^A) dx^A dx^B 
\end{equation}
then the infinitesimal transformations 
$\xi^t =$ constant, $\xi^x =a x f(x^A), \xi^A = q^A(x,x^A)$,
subject to the conditions $q^A (0,x^A) =0$ and
$\partial_x q^A = - \sigma^{AB} \partial_B f(x^A)$,
leave the induced metric on the horizon invariant.
(This is quite similar to the transformations in the case of AdS;
see for example, \cite{adscft}.]
In the same vein, one can construct the metric in the bulk in a Taylor
series expansion in $x$, from the form of the metric near the horizon,
along the lines of exercise 1 (page 290) of \cite{ll}.
These ideas work only because, algebraically, $N\to 0$ makes certain terms
in the diffeomorphisms vanish and increases the symmetry.
There is a strong indication that most of the results related to horizons
(such as entropy) will arise from the enhanced symmetry of the theory near
the $N=0$ surface (see e.g. \cite{carlip} and references therein).

(f) Space-time gauges with $N=0$ hypersurfaces also create difficulties in
canonical quantum gravity using, say, Wheeler-DeWitt equation.
The solution to the Wheeler-DeWitt equation $\Psi [{}^3\mathcal{G}]$
is a functional of the 3-geometry ${}^3\mathcal{G}$,
which is defined to be the equivalence class of 3-metrics $g_{\alpha\beta}$
connected by purely spatial 3-dimensional coordinate transformations.
$N$ exists only as a gauge variable in the theory,
and is usually integrated out in determining, for example,
the path integral kernel between two 3-geometries.
However, the entire analysis breaks down if $N\to 0$,
and it is not clear what is the role of the large gauge transformations
in the full theory of quantum gravity.
A careful analysis \cite{teit} leads to the conclusion:
``In the classical theory one may perform changes of the four space-time
coordinates and the action remains invariant.
However, in the final equations of the quantum theory,
there is only room for the changes of the spatial coordinates \ldots
but there is no place for reparametrizations of the time coordinate."
It is therefore not clear how the large gauge transformations,
such as the one from Minkowski metric to Rindler metric,
are incorporated in the full quantum gravity.
In the same analysis \cite{teit},
it was noted that one needs to make a choice for $N$ to perform computations,
and ``\ldots $N$ must be different from zero at all times".
These issues have surfaced and have been dealt with at different levels
of rigor and correctness in several other works dealing with semiclassical
gravity, but a study of literature shows that it is not easy to incorporate
$N \to 0$ gauges in the analysis of semiclassical gravity.
In our opinion, this is because the standard approach attempts to provide a
\emph{global} theory of quantum/semiclassical gravity, which cannot exist in
a gauge in which a class of observers have only limited access to space-time.
It is necessary to take in to account the role of large gauge transformations
separately, which is what we do in this paper. 
  
(g) The existence of gauges in which $N=0$ also has important implications
for Euclidean field theory, which is widely used to ``define" quantum field
theoretic expressions.
The simple procedure of analytic continuation in time coordinate
$t\to \tau=it$ is not generally covariant and does not ``commute"
with general coordinate transformations.
A well-known example is that of Rindler frame in (\ref{stdrindler}),
where the quantum theories defined by analytic continuation in $t$ and in $T$
are widely different; in fact it can be shown that there is no unitary
transformation connecting these theories \cite{gerlach}.
Either we should abandon these gauges as not unitarily implementable
in quantum theory, or accept the fact that each of these gauges will define
for a class of observers their own quantum description.
The former approach runs in to serious problems in the classical limit,
and it is the latter approach which we take in this paper.

In summary, we take the singular gauges in which $N=0$ on a hypersurface
(with other metric components remaining well behaved) seriously.
The usual attitude is to claim that ``this is only a relabelling of
coordinates" and ``physics should not change under such relabelling".
But these singular gauges allow for class of observers
with limited access to space-times,
and when we demand that any such observer has a right to
formulate physics in terms of variables he can access,
new constraints on the dynamics emerge.
This approach to combining general covariance and quantum theory leads
to as far reaching conclusions as the combination of Lorentz covariance
and quantum theory.

\section{The Action Principles for Gravity}

The  arguments given at the end of section I show the power of introducing the principle of
effective theory.
Needless to say, only a very special kind of action $A_{\rm grav}$ can fit
in to the above described structure naturally.
Incredibly enough, the conventional action principle used for gravity fits
the bill, though it was never introduced in this light.
We shall now describe how this comes about.
(The approach presented here starts from the Einstein-Hilbert action, which
 can be obtained from considerations of general covariance.
Alternatively, it is possible to  start from the form of the boundary term
and invoke general covariance to obtain the Einstein-Hilbert action, which
was  explored in
\cite{tpwork2,tpwork3}.)

To the lowest non-trivial order, the effective action for gravity
comprises of terms containing up to two derivative operators.
The Einstein field equations of gravity are generally covariant
and of second order.
Taking the metric tensor $g_{ab}$ as the fundamental variable,
one would naively expect these equations to be derived from an action
principle involving $g_{ab}$ and its first derivatives $\partial_k g_{ab}$,
analogous to the situation for many other field theories of physics.
But it is not possible to construct a generally covariant action
for gravity out of only $g_{ab}$ and $\partial_k g_{ab}$.
It is, however, possible to obtain the field equations using a generally
covariant Einstein-Hilbert action which contains second derivatives
of the metric tensor: 
\begin{equation}
A_{\rm EH} \equiv {1\over 16\pi} \int d^4x~\sqrt{-g}~R ~.
\end{equation}  
In spite of this action containing second derivatives of the metric tensor,
the field equations obtained from it are only of second order.
This happens because the second derivative part of $A_{\rm EH}$ can be
separated as a total divergence;
neither this total divergence term nor the remaining first derivative
part is generally covariant.
(The most general effective action for gravity also contains a
cosmological constant term, obtained by adding a real constant to $R$.
This term is not important in our analysis, and so we leave it out.)

To see the structure of the terms more clearly, we introduce a
$(1+3)$ foliation with the standard notation for the metric components
($g_{00}=- N^2,~g_{0\alpha}=N_\alpha$).
Let $u^i=(N^{-1},0,0,0)$ be the four-velocity of observers
corresponding to this foliation, i.e. the normal to the foliation,
and let $a^i=u^j\nabla_ju^i$ be the related acceleration.
Let $K_{ab}=-\nabla_a u_b - u_aa_b$ be the extrinsic curvature
of the foliation, with $K\equiv K^i_i = -\nabla_i u^i$.
(With this standard definition, $K_{ab}$ is purely spatial,
$K_{ab}u^a=K_{ab}u^b=0$; so one can work with the spatial components
$K_{\alpha\beta}$ whenever convenient.)
Then it is easy to show that (this result is mentioned in ref.\cite{mtw},
p.520, eq.(21.88); a simple derivation is given in the Appendix A)
\begin{equation}
R \equiv L_{\rm EH}
= L_{\rm ADM} - 2\nabla_i(Ku^i + a^i) \equiv L_{\rm ADM}+L_{\rm div}
\label{ehandadm}
\end{equation}
where 
\begin{equation}
L_{\rm ADM} = {}^{(3)}{\cal R} + (K_{ab}K^{ab} - K^2)
\label{defadml}
\end{equation}
is the ADM Lagrangian \cite{ADM} quadratic in $\dot g_{\alpha\beta}$,
and $L_{\rm div} = -2\nabla_i(Ku^i + a^i)$ is a total divergence.
Neither $L_{\rm ADM}$ nor $L_{\rm div}$ is generally covariant.
For example, $u^i$ explicitly depends on $N$,
which changes when one makes a coordinate transformation
from the synchronous frame to a frame with $N\neq 1$.

There is a conceptual difference between the $\nabla_i(Ku^i)$ term
and the $\nabla_i a^i$ term that occur in $L_{\rm div}$.
This is obvious in the standard foliation, where $Ku^i$ contributes
on the constant time hypersurfaces, while $a^i$ contributes on the
time-like or null surface which separates the space in to two regions
(as in the case of a horizon).
To take care of the $Ku^i$ term more formally, we recall that
the form of the Lagrangian used in functional integrals depends
on the nature of the transition amplitude one is interested in computing,
and one is free to choose a suitable perspective.
For example, in non-relativistic quantum mechanics, if one uses the
coordinate representation, the probability amplitude for the dynamical
variables to change from $q_1 $ (at $t_1$) to $q_2$ (at $t_2$) is given by
\begin{equation}
\psi(q_2,t_2) = \int dq_1 K \left( q_2,t_2;q_1,t_1 \right) \psi(q_1,t_1) ~,
\end{equation}
\begin{equation}
K \left( q_2,t_2;q_1,t_1 \right) = \sum\limits_{\rm paths}
\exp \left[ {i\over\hbar} \int dt~L_q(q,\dot q) \right] ~,
\label{qsopa}
\end{equation}
where the sum is over all paths connecting $(q_1,t_1)$ and $(q_2,t_2)$,
and the Lagrangian $L_q(q,\dot q)$ depends on $(q,\dot q)$. 
It is, however, quite possible to study the same system in momentum space,
and enquire about the amplitude for the system
to have a momentum $p_1$ at $t_1$ and $p_2$ at $t_2$. 
From the standard rules of quantum theory, the amplitude for the particle
to go from $(p_1,t_1)$ to $(p_2,t_2)$ is given by the Fourier transform
\begin{equation}
G \left( p_2,t_2;p_1,t_1 \right) \equiv \int dq_2 dq_1
~K \left( q_2,t_2;q_1,t_1 \right)  
~\exp \left[ -{i\over\hbar} \left( p_2 q_2 - p_1 q_1 \right) \right] ~.
\label{qftofq} 
\end{equation}
Using (\ref{qsopa}) in (\ref{qftofq}), we get
\begin{eqnarray}
G \left( p_2,t_2;p_1,t_1 \right) &=& \sum\limits_{\rm paths}
\int dq_1 dq_2 \exp \left[ {i\over\hbar}
\left\{ \int dt~L_q - \left( p_2 q_2 - p_1 q_1 \right) \right\} \right]
\nonumber \\
&=& \sum\limits_{\rm paths} \int dq_1 dq_2 \exp \left[ {i\over\hbar}
\int dt \left\{ L_q - {d \over dt} \left( pq \right) \right \} \right]
\nonumber \\
&=& \sum_{\rm paths} {} \exp \left[ {i\over\hbar}
\int L_p(q, \dot q, \ddot q)~dt \right] ~.
\label{lp}
\end{eqnarray}
In arriving at the last expression, we have
(i) redefined the sum over paths to include integration over $q_1$ and $q_2$;
and (ii) upgraded the status of $p$ from the role of a parameter in the
Fourier transform to the physical momentum $p(t)=\partial L/\partial \dot q$.
This result shows that, given any Lagrangian $L_q(q,\partial q)$
involving only up to the first derivatives of the dynamical variables,
it is \emph{always} possible to construct another Lagrangian
$L_p(q,\partial q,\partial^2q)$ involving up to second derivatives,
such that it describes the same dynamics but with different boundary
conditions \cite{tpwork2}.
The prescription is:
\begin{equation}
L_p = L_q - {d\over dt} \left( q{\partial L_q \over \partial\dot q} \right) ~.
\label{lbtp}
\end{equation}
While using $L_p$, one keeps the \emph{momenta} $p'$s fixed
at the endpoints rather than the \emph{coordinates} $q'$s.
This boundary condition is specified by the subscripts on the Lagrangians.

The result generalizes directly to multi-component fields.
If $q_A(x^i)$ denotes a component of a field (which could be a component
of a metric tensor $g_{ab}$, with $A$ formally denoting pairs of indices),
then we just need to sum over $A$.
Since $L_{ADM}$ is quadratic in $\dot g_{\alpha\beta}$, we can treat
$g_{\alpha\beta}$ as coordinates and obtain another Lagrangian $L_\pi$
in the momentum representation.
The canonical momentum corresponding to $q_A=g_{\alpha\beta}$ is
\begin{equation}
p^A = \pi^{\alpha\beta}
= \frac{\partial (\sqrt{-g}~L_{ADM})}{\partial \dot g_{\alpha\beta}}
=- \sqrt{-g} {1 \over N} (K^{\alpha\beta}- g^{\alpha\beta}K) ~,
\end{equation}
so that the term $d(q_Ap^A)/dt$ is just the time derivative of
\begin{equation}
g_{\alpha\beta}\pi^{\alpha\beta}
= -\sqrt{-g} {1 \over N}(K-3K) = \sqrt{-g} (2Ku^0) ~.
\end{equation}
Since
\begin{equation}
{\partial \over \partial t} (\sqrt{-g}~Ku^0)
= \partial_i (\sqrt{-g}~Ku^i) = \sqrt{-g}~\nabla_i (Ku^i) ~,
\end{equation}
the combination
$\sqrt{-g}~L_\pi \equiv \sqrt{-g} [L_{\rm ADM} - 2 \nabla_i(Ku^i)]$
describes the same system in the momentum representation with
$\pi^{\alpha\beta}$ held fixed at the end points \cite{york}.
Switching over to this momentum representation, the relation between the
action functionals corresponding to (\ref{ehandadm}) can now be expressed as 
\begin{equation}
A_{\rm EH} = A_\pi + A_{\rm boundary} ~,
\end{equation}
\begin{equation}
A_\pi \equiv A_{\rm ADM} - {1 \over 8\pi}\int\sqrt{-g}~d^4x~\nabla_i(Ku^i) ~.
\label{defmomspace}
\end{equation}
Here $A_\pi$ describes the ADM action in the momentum representation, and 
\begin{equation}
A_{\rm boundary} =- {1 \over 8\pi} \int d^4x~\sqrt{-g}~\nabla_ia^i
= -{1 \over 8\pi} \int dt \int_{\cal S} d^2x~N \sqrt{\sigma}
(n_\alpha a^\alpha) 
\label{boundary}
\end{equation} 
is the boundary term arising from the integral over the surface.
In the last equality,
$\sigma_{\alpha\beta} = g_{\alpha\beta} - n_\alpha n_\beta$
is the induced metric on the boundary 2-surface with outward normal
$n_\alpha$, and the gauge $N_\alpha=0$ has been chosen. 

\section {Semiclassical Quantization of Gravity}

In case of space-times without boundary, it does not matter if one works
with $A_{\rm EH}$ or $A_{\rm ADM}$ or $A_\pi$, since they all differ from
each other by total divergences.
On the contrary, while dealing with space-times with boundaries,
it is crucial to use an appropriate action in the functional integral
for gravity.
We believe that the correct action to use in the functional integral is
$A_{\rm ADM}$ or equivalently $A_\pi$ (which describes the same system
in the momentum representation), since it is quadratic
in the time derivatives of the true dynamical variables $g_{\alpha\beta}$. 

To study the effects of unobserved degrees of freedom in some space-time
region, let us divide the space-time manifold in to two regions with a
boundary surface separating them, and choose a coordinate system such that
this boundary acts as a horizon for the observer on one side (side 1, say).
The effective theory for this observer is obtained by integrating out the
variables on the inaccessible side (side 2):
\begin{equation}
\exp [iA_{\rm eff}(g_1)] \equiv \int [{\cal D} g_2]
~\exp [i(A_{\pi} (g_1) + A_{\pi}(g_2))] ~.
\end{equation} 
We have chosen $A_\pi$ to be the action functional describing gravity,
so the functional integral is to be evaluated holding the extrinsic
curvature of the boundary fixed.
In the absence of any matter, we have $R=0$ for the classical solution of gravity.
It follows that $A_{\rm EH}^{\rm WKB}=0$, and
\begin{equation}
A_{\pi} (g_2^{\rm WKB}) = - A_{\rm boundary}^{n_2}(g_2)
= A_{\rm boundary}^{n_1}(g_1)
= -{1 \over 8\pi} \int d^4x~\sqrt{-g}~\nabla_ia^i ~,
\label{divdef}
\end{equation}
where $n_1=-n_2$ denote the outward normals of the two sides.
(As an aside, we mention that if the matter on side 2 is described by a
scale invariant action, making energy-momentum tensor traceless, $T=0$,
then the same result holds.
If the matter has non-zero $T$, then we get an extra phase factor involving
the volume integral of $T$ but independent of the boundary degrees of freedom.
This phase factor does not affect our conclusions, since we are only
concerned with phases which \emph{change} under co-ordinate transformations.
We are not including matter degrees of freedom in our discussion here
and hope to address them in a future publication.)
Thus, in the semiclassical limit, we indeed obtain a boundary term
involving the gravitational degrees of freedom,
as anticipated by the principle of effective theory. 
Using this result, the effective theory on side 1 is described by the action 
$A_{\rm eff}^{\rm WKB}(g_1)$ with 
\begin{equation}
\exp [iA_{\rm eff}^{\rm WKB}(g_1)]
= \exp [i(A_{\pi}(g_1) + A_{\rm boundary}^{n_1} (g_1))] \times {\rm det}(Q) ~,
\end{equation}
where det$(Q)$ arises from integration over quantum fluctuations,
and can be ignored in the lowest order analysis.
(Incidentally, we mentioned earlier --- see equation (\ref{scalar}) ---
 that the gravitational action producing
only a boundary term in the semiclassical limit is highly nontrivial,
and gave the counter-example of the scalar field theory in coordinate
representation. Changing to momentum representation does not help in the
case of a scalar field. The action for an interacting scalar field cannot 
be reduced to a pure boundary term even in the  momentum representation.)

The extra term, $A_{\rm boundary}$, is a total divergence and does not
change the equations of motion for side 1 in the classical limit.
But it can affect the quantum theory, unless
$2\sqrt{-g}~\nabla_ia^i = 2 \partial_\alpha (\sqrt{-g}~a^\alpha)$
does not contribute. In certain situations this term vanishes:
(a) If one uses a synchronous coordinate system with $N=1,~N_{\alpha}=0$,
in which there is no horizon.
In the case of a black hole space-time, for example, this coordinate system
will be used by a class of in-falling observers.
(b) If the integration limits for $2 \partial_\alpha (\sqrt{-g}~a^\alpha)$
could be taken at, say origin and spatial infinity, where the contribution
actually vanishes due to $N_\alpha\to 0,~N\to$ constant at these limits.

For a generic observer, however, we cannot ignore the contribution of this
term, and we have to deal with the boundary action (\ref{boundary}).
More explicitly, if we compare the synchronous frame
(for which $N=1,~N_{\alpha}=0$), with the one obtained by the infinitesimal
transformation in (\ref{largegauge}) (for which $N\ne1,~N_{\alpha}=0$),
we note that the value of the boundary term can change.
Our principle of effective theory requires that this coordinate/observer
dependent term should only covariantly affect the quantum amplitudes.
The only way to ensure this is to make $\exp[iA_{\rm boundary}]$
single-valued, i.e. demand that 
\begin{equation}
A_{\rm boundary} = 2\pi n + {\rm constant},~ n={\rm integer}.
\end{equation}
Then $\exp[iA_{\rm boundary}]$ becomes an overall phase, and the physics
on side 1 is determined by $A_\pi(g_1)$ as originally postulated.
The values of $A_{\rm boundary}$ measured from side 1 and side 2 are of
opposite sign, because of the opposite direction of their outward normals.
The spectrum of $A_{\rm boundary}$ is therefore symmetric about zero:
\begin{equation}
A_{\rm boundary} = 2\pi m ~,
\label{spectrum}
\end{equation}
with two possible sequences for $m$, either $m \in \{0,\pm1,\pm2,\ldots\}$
or $m \in \{\pm1/2,\pm3/2,\ldots\}$.
The boundary term---which is not generally covariant---may be different
for different observers, but the corresponding quantum operators need
not commute, thereby eliminating any possible contradiction.  
(This is analogous to the fact that, in quantum mechanics,
the component of angular momentum measured along any axis
is quantized irrespective of the orientation of the axis.)
The action with a uniformly spaced spectrum, $A=2\pi m\hbar$,
has a long and respectable history in quantum field theories,
and our analysis gives a well defined realization of this property
for the semiclassical limit of quantum gravity.

In Lorentzian space-time, there are no natural limits on the time
integration in (\ref{boundary}), and the numerical value of
$A_{\rm boundary}$ depends on the range chosen for the integration.
When analytical continuation to Euclidean time is carried out
in a coordinate system with a horizon, the time coordinate \emph{must}
be made periodic to avoid a conical singularity.
Thus, in space-times with horizons, there is a natural periodicity
requirement on the \emph{Euclidean} time coordinate, with period $\beta$.
In such space-times, we take the range of time integration to be $[0,\beta]$.
With this assumption, the quantization condition for space-times with
horizons becomes 
\begin{equation}
A_{\rm boundary} = -{1 \over 8\pi} \int_0^\beta dt
\int d^2x~N \sqrt{\sigma}~(n_\alpha a^\alpha) = 2\pi m ~,
\label{condition}
\end {equation} 
with $m\ge0$ for the observer outside the one-way horizon.
This result has several important consequences \cite{comment}:

(i) In all static space-times with horizons, it can be shown that
this boundary term is proportional to the area of the horizon.
As the surface approaches the one-way horizon from outside, the quantity
$N(a_in^i)$ tends to $(-\kappa)$, where $\kappa$ is the surface gravity
of the horizon and is constant over the horizon \cite{surfacegrav}.
Using $\beta\kappa=2\pi$, the contribution of the horizon becomes
\begin{equation}
A_{\rm boundary}=\frac{1}{4}({\rm Horizon~Area})
\equiv\frac{1}{4}A_{\rm horizon}
\label{contribution}
\end {equation}
Our result therefore implies that the area of the horizon,
as measured by \emph{any} observer blocked by that horizon, will be quantized.
(In normal units, $A_{\rm boundary}=2\pi m\hbar$ and
$A_{\rm horizon}=8\pi m(G\hbar/c^3)=8\pi mL_{\rm Planck}^2$).
In particular, any flat spatial surface in Minkowski space-time can be made
a horizon for a suitable Rindler observer, and hence all area elements
in even flat space-time must be intrinsically quantized.
In the quantum theory, the area operator for one observer
need not commute with the area operator of another observer,
and there is no inconsistency in all observers measuring quantized areas.
The changes in area, as measured by any observer, are also quantized,
and the minimum detectable change is of the order of $L_{\rm Planck}^2$.
It can be shown, from very general considerations, that there is an
operational limitation in measuring areas smaller than $L_{\rm Planck}^2$,
when the principles of quantum theory and gravity are combined \cite{tplimit};
our result is consistent with this general analysis.

While there is considerable amount of literature  suggesting that the area of \emph{a black hole
  horizon} is quantized (for a small sample of references, see \cite{areaquant} )
  we are not aware of any result which is as general as suggested above or derived
  so simply. Even in the case of a Schwarzschild black hole all the results in the literature
  do not match in detail, nor is it conceptually easy to relate them to one another.
  For example, a simple procedure to derive the area spectrum of a Schwarzschild
  black hole from canonical quantum gravity is to proceed as follows. 
  Classically, the outside region of any spherically symmetric collapsing matter of finite
  support is described by the Schwarzschild metric with a single parameter $M$.
  Since the pressure vanishes on the surface of the collapsing matter,
any particle located on the surface will follow a time-like geodesic
trajectory $a(t)$ in the Schwarzschild space-time.
   Because of the extreme symmetry of the model as
  well as the constancy of $M$, the dynamics of the system  can be mapped to the
  trajectory of this particle. The action describing this trajectory 
  can be taken to be 
  \begin{equation}
  A = \int dt~(p\dot{a} - H(p,a)) ~,~~
H = \frac{1}{2} \left( \frac{p^2}{a} + a \right) ~,
  \label{hamqc}
  \end{equation}
  which is precisely the action that arises in the study of closed dust-filled Friedmann models
  in quantum cosmology. (It is possible to introduce a canonical transformation from $(p,a)$ to
  another set of variables $(P,M)$ such that the Lagrangian becomes $L= P \dot M - M$.
  This is similar to the set of variables introduced by Kuchar \cite{kuchar} in his analysis of spherically symmetric
  space-times.)
  This is expected since --- classically --- 
   the Schwarzschild exterior can be matched to a homogeneous
  collapsing dust ball, which is just the Friedmann universe as the 
  interior solution. Now, it is obvious that
  there is {\it no} unique quantum theory for the Hamiltonian in (\ref{hamqc}) because of  operator ordering problems. However, several sensible
  ways of constructing a quantum theory from this Hamiltonian lead to discrete area spectrum
  as well as a lower bound to the area (see for example \cite{tpqc}). 
  There are many variations on this theme as far as the area of black hole
horizon is concerned, but we believe our approach is conceptually simpler
and bypasses many of the problems faced in other analyses.

(ii) The boundary term originated from our integration of the unobserved
gravitational degrees of freedom hidden by the horizon.
Such an integration should naturally lead to the entropy of the unobserved
region, and we get---\emph{as a result}---that the entropy of a horizon is
always one quarter of its area.
Our analysis also clarifies that this horizon entropy is the contribution
of quantum entanglement across the horizon
of the gravitational degrees of freedom.
Further it makes \emph{no} distinction between different types of horizons,
e.g. the Rindler and Schwarzschild horizons.
In contrast to earlier works, most of the recent works---especially the ones
based on CFT near horizon---do not make any distinction between different
types of horizons.
We believe all horizons contribute an entropy proportional to area for
observers whose vision is limited by those horizons.
As we stressed before, the entropy of the black hole is also observer dependent,
and freely falling observers plunging in to the black hole will not attribute
any entropy to the black hole.

(iii) More generally, the analysis suggests a remarkably simple,
thermodynamical interpretation of semiclassical gravity.
In any static space-time with the metric
\begin{equation}
ds^2 = -N^2({\bf x})~dt^2 + \gamma_{\alpha\beta}({\bf x})~dx^\alpha dx^\beta ~,
\end{equation}
we have $R = {}^{(3)}{\cal R}-2\nabla_ia^i$,
where $a_i=(0,\partial_\alpha N/N)$
is the acceleration of ${\bf x}=$ constant world-lines.
Then, limiting the time integration to $[0,\beta]$,
the Einstein-Hilbert action becomes
\begin{equation}
A_{\rm EH} = {\beta \over 16\pi} \int_{\cal{V}} d^3x
~N\sqrt{\gamma}~{}^{(3)}{\cal R} - {\beta \over 8\pi} \int_{\partial\cal{V}}
d^2{\cal S}~N(n_\alpha a^\alpha) \equiv \beta E - S ~,
\end{equation}
In the Euclidean sector, the first term is proportional to energy
(in the sense of the spatial integral of the ADM Hamiltonian),
and the second term is proportional to entropy in the presence of a horizon.
$A_{\rm EH}$ thus represents the free energy of the space-time,
and various thermodynamic identities follow from its variation.
(This equivalence is explored in detail for spherically symmetric
space-times in \cite{tpcqg}).

(iv) The boundary term can be given a topological meaning
in Euclidean coordinates \cite{tpinstanton}.
Near the horizon, one can expand the metric coefficients in a Taylor series
and approximate the metric, in suitable coordinates, to the Rindler form:
\begin{equation}
ds^2 \approx -(ax)^2 dt^2 + dx^2 + dL_\perp^2
\label{rindler}
\end{equation}
This coordinate system is related to the local inertial frame by
$(X=x \cosh at,~T=x \sinh at)$, with the curves of constant $x$ being
the hyperbolas $X^2-T^2=x^2$ in the local inertial frame.
If we analytically continue the inertial time coordinate, $T\to -iT_E$,
these hyperbolas become circles around the origin $X^2+T_E^2=x^2$,
and the horizon (corresponding to $x=0$ in Rindler coordinates and $X=\pm T$
in the inertial coordinates) becomes the \emph{single point} at the origin.
The surface $x=\epsilon\to 0$, infinitesimally close to the horizon for a
Rindler observer, becomes a circle of infinitesimal radius around the origin.
The boundary contribution from the horizon can be interpreted in terms of
a topological winding number around the origin \cite{chrisduff}.
In fact, this result is of a very general validity, since one can construct
a local inertial frame and a local Rindler frame around any event.
This result also shows that the large gauge transformations,
which make $N$ vanish along a surface, 
can have nontrivial effects in the Euclidean sector,
since   analytic continuation in the time coordinate
is not a generally covariant procedure.
Effectively, this large gauge transformation leads to ``punctures"
in the Euclidean sector and to winding numbers \cite{inprogress}.

\section {First law of Horizon thermodynamics and some further generalizations}

The crucial feature which we have exploited is that the conventional action
for gravity contains a boundary term involving the integral of the normal
component of the acceleration.
As far as we know, this term has not been brought to center stage in any of the previous analyses. 
Since the action for matter and gravity will be additive in the full theory, 
the boundary term also has implications for the manner in which the dynamics of the 
horizon will change in the full theory. This in turn will require handling a horizon  which is
 time dependent,  in the sense that $N=0$ on the surface
$x=x_H(t)$ in some suitably chosen coordinate system. We shall briefly comment on these issues though a complete discussion is postponed to a future publication.

Our approach can  be generalized along the following lines 
to a more general context in which the horizon is time dependent. 
  Though there are few realistic solutions with time dependent horizons, it is possible to model
this situation using a generalization of Rindler frame for time dependent acceleration given in (\ref{eqn:fiftyfive}). The metric in (\ref{eqn:fiftyseven})
provides a good approximation
to the space-time close to any time dependent horizon.
In this case, the action is entirely a surface term,
\begin{equation}
 \frac{dA}{d\mathcal{A}_\perp}= \frac{1}{8\pi} \int g(t) dt =   \frac{1}{8\pi} (\mu_2 -\mu_1)~,
\label{timedeprind}
\end{equation}
where $\mathcal{A}_\perp$ is the transverse proper area (in the $y-z$ plane).
From the coordinate transformations in equation (\ref{eqn:fiftyfive}),
it is obvious that the Euclidean continuation will require
$t \to it,~\mu \to i\mu$ as well as some conditions on the
functional form of $g(t)$ to ensure reality of the Euclidean metric.
Hence we expect $\mu$ to be periodic with a period $2\pi$ in the 
Euclidean sector. This corresponds to the condition 
\begin{equation}
\int g(t) dt = 2\pi \label{cond} ~,
\end{equation}
for the absence of conical singularities in the Euclidean sector.
Given this condition, we again find that the contribution to the action
is one quarter of the transverse horizon area.
This result is applicable even for $g(t)$ which starts with $g=0$
(with space-time represented in inertial coordinates) for $t<t_0$,
and evolves to a space-time with a horizon asymptotically.
Such a situation models the collapse of a system to form a black hole
with a horizon.

The above result can be generalized in a manner which throws light
on another aspect of our analysis. 
Our result that the semi-classical action is quantized with
$A_{\rm boundary}=2\pi m$ is very reminiscent of the 
``old" quantum theory in which one often uses the condition 
\begin{equation}
\oint p_a dx^a \approx 2\pi n
\end{equation}
Though our result is not in the above form,
it can be reinterpreted in a manner which brings in the above connection.
To achieve this, we begin by studying the variation of the semiclassical
action, $\delta A$, when the metric is varied by $\delta g_{ab}$.
In case of the quadratic action (see Appendix A for details of the notation),
\begin{equation}
A \equiv \frac{1}{16\pi}\int_\mathcal{V} d^4x \sqrt{-g}~R
  +\frac{1}{8\pi}\int_{\partial\mathcal{V}} d^3x \sqrt{{}^{(3)}g}~K ~,
\end{equation}
the variation $\delta A=0$ if $\delta g_{ab}=0$ on $\partial\mathcal{V}$
and the equations of motion are satisfied.
Hence the the only contribution to $\delta A$ arises from the variation
of the metric on the boundary and is given by 
\begin{equation}
\delta A = \frac{1}{16\pi} \int_{\partial\mathcal{V}} d^3x \sqrt{|f|}~
           (Q^{ab} - f^{ab} Q)~\delta f_{ab} ~,
\label{actionvariation}
\end{equation}
where $f_{ab}$ is the induced metric on the boundary $\partial \mathcal{V}$
and $Q_{ab}$ is the extrinsic curvature of the surface.
To be specific, consider a four volume ${\cal V}$ bounded by two space-like
hypersurfaces $\Sigma_1$ and $\Sigma_2$ and a time-like hypersurface ${\cal S}$
(which will become a horizon in the limiting case), as described in Appendix A.
Let us now consider the variation of the metric induced by relabelling of
coordinates on the boundary. If the coordinates on the boundary are shifted
infinitesimally, $x^a\to x^a+\xi^a(x)$, the resulting change in the metric
is of the form $\delta f_{ab} = D_{(a} \xi_{b)}$
where $D_a$ is the covariant derivative operator defined on the boundary.
We introduce this form for $\delta f_{ab}$ in to (\ref{actionvariation})
and integrate by parts.
One of the constraint equations of general relativity requires
$D_a (Q^{ab} - f^{ab} Q) =0$, allowing the integration to be converted
to a surface integral.
Since we are interested in the horizon, we shall take $\partial \mathcal{V}$
to be the time-like surface $\mathcal{S}$.
Then the surface integral picks up a contribution on $\mathcal{Q}$,
\begin{equation}
\delta A = \frac{1}{8\pi} \int_{\mathcal{Q}} d^2 x \sqrt{\sigma } \xi^b w^a 
(\Theta_{ab} - \gamma_{ab} \Theta) ~,
\end{equation}
where $w^a$ is the normal to $\mathcal{Q}$ when it is treated as a surface
embedded in $\Sigma$.
This normal is the same as $u^a =(N^{-1},0,0,0)$, leading to 
\begin{equation}
\frac{dA}{\sqrt{\sigma} d^2 x}= \frac{1}{8\pi} \xi^b u^a 
(\Theta_{ab} - \gamma_{ab} \Theta) ~,
\end{equation}
which gives the change in contribution to the action per unit transverse
proper area induced by the coordinate change.
(The usual result that the action is invariant under coordinate transformations
is based on the assumption that $\xi^a$ vanishes on $\partial\mathcal{V}$;
we are interested in cases where this is not true.)
Noting that $\xi^b$ is infinitesimal, we can convert it in to an integration
measure by setting $\xi^b=(1/2)dx^b$. (The factor (1/2) takes in to account
the symmetrization condition on $\delta g_{ab}$.)
Integrating over this measure, we get 
\begin{equation}
\frac{dA}{\sqrt{\sigma} d^2 x}= \frac{1}{16\pi} \int dx^b u^a 
(\Theta_{ab} - \gamma_{ab} \Theta) \equiv \oint p_b dx^b ~,
\label{pdq}
\end{equation}
which gives the contribution to the action per unit transverse proper area
as an integral of $p dq$.
It must be stressed that no adiabatic approximation is made in arriving at
this result, and it is exact when interpreted along the lines indicated.
(The variables $p_a$ can be interpreted---in a limited sense---as momenta
associated with the surface degrees of freedom of gravity; but $x^a$ are
just the ordinary coordinates and not the dynamical variables of gravity.)
  
In case of a horizon, the most important shift corresponds to $x^b=x^0$.
A simple calculation shows that
$\Theta_{00} = -N^2 (a_\alpha n^\alpha) ,~ \Theta = q - a_\alpha n^\alpha$,
giving 
\begin{equation}
\frac{dA}{\sqrt{\sigma}~d^2x}
= \frac{1}{16\pi} \int dt (2N a_\alpha n^\alpha - N q) ~.
\end{equation}
As we approach the horizon, $N a_\alpha n^\alpha \to -\kappa$ and $Nq \to 0$.
Then the contribution to the action per unit transverse proper area is given by
\begin{equation}
\frac{dA}{\sqrt{\sigma} d^2 x}=- \frac{1}{8\pi} \int \kappa dt ~.
\label{actchange}
\end{equation}
This result is applicable even for time dependent $\kappa$,
and the absence of conical singularity in the Euclidean sector requires
$\int \kappa dt = 2\pi$. 
Thus we obtain the previous result that the contribution to the action
is one quarter per unit transverse proper area.
The result (\ref{actchange}) matches with (\ref{timedeprind})
since both can be interpreted as due to changes in the coordinates.

More generally, (\ref{actchange}) can be written in the form
\begin{equation}
\delta A = -\frac{\delta \mathcal{A}_\perp}{8\pi} (\mu_2 -\mu_1) ~,
\end{equation}
where $\mu$ measures the hyperbolic angle which becomes a trigonometric
angle in the Euclidean sector with $d\mu/dt =\kappa$.
A variant of this equation has occurred in the literature before
(see e.g. \cite{carlipteit}) in the form of a Schrodinger equation
$-i\hbar (\partial \psi /\partial \mu) = \mathcal{A}_\perp \psi$, 
in which $\mu$ is treated as a dimensionless internal time conjugate
to the area operator.
When the horizon is in interaction with matter fields,
the change of phase of the wave function in the gravity sector is 
\begin{equation}
\delta A=
-\frac{\kappa\delta \mathcal{A}_\perp}{8\pi} (t_2 - t_1)=
-\frac{\kappa}{2\pi} \frac{(\delta \mathcal{A}_\perp)}{4}(t_2 - t_1) ~,
\end{equation}
whereas a change of energy of $\delta E$ should lead to a phase change of
$-\delta E (t_2-t_1)$. This allows us to identify 
\begin{equation}
\delta E=\frac{\kappa}{2\pi} \frac{(\delta \mathcal{A}_\perp)}{4}=T\delta S ~,
\end{equation}
ensuring consistency with first law of horizon thermodynamics.
More generally, the semiclassical limit of Wheeler-DeWitt equation will have the classical action for 
 gravity in its phase, which in turn will determine the time coordinate for matter evolution \cite{semiclassical}. 
 If energy is exchanged between matter and gravity this will lead to phase changes,
and the requirement of consistency will lead to the determination of the boundary term of the 
gravitational action. It can be shown \cite{tpwork2,tpwork3} that this is enough to determine the full 
form of the action for gravity. Thus the low energy description of gravity is tightly constrained by the 
horizon dynamics.

\section{Summary and Outlook}

Gravity is a geometric theory of space-time, and its natural setting is in
terms of the metric tensor as a function of the coordinates, $g_{ab}(x)$.
The fact that there is a maximum speed for propagation of any physical
signal, i.e. the speed of light, means that no observer can access phenomena
occurring outside his/her light cone.
There exist choices of space-time coordinates, where the global light cone
structure makes part of the space-time inaccessible to a class of observers.
A realistic physical theory must be formulated in terms of whatever
variables the observer has access to;
contribution of the unobservable regions must be such that it can be
re-expressed in terms of the accessible variables.
This principle of effective theory provides a powerful constraint on the
theory of gravity, when it is demanded that the same formulation of the
theory should be used by all observers, irrespective of whether or not
their coordinate choice blocks their access to some regions of space-time.

To understand the significance of this demand, it is instructive to look back
at how the special theory of relativity is combined with quantum dynamics.
Non-relativistic theories have an absolute time and exhibit invariance under
the Galilean group.
In the path integral representation of non-relativistic quantum mechanics,
one uses only the causal paths $x^\alpha(t)$ which ``go forward"
in this absolute time coordinate.
This restriction has to be lifted in special relativity and the corresponding
path integrals use paths $x^a(s)=(t(s),x^\alpha(s))$, which go forward
in the proper-time $s$ but either forward or backwards in coordinate time $t$.
Although propagation of real particles is restricted to be within
their light cones, virtual particles can follow a space-like trajectory
and propagate outside their light cones.
To recover a causal theory (i.e. the degrees of freedom outside an observer's
light cone should not influence his/her evolution), one must introduce the
notion of a quantum field, antiparticles and $i\epsilon$-prescription for
retarded propagators.
In this special relativistic formulation,
the allowed transformations are those of the Lorentz group.
In particular, the light cone structure is invariant under Lorentz boosts,
and hence all observers at the same space-time point see the same
light cone structure.
The situation becomes more complicated in the general theory of relativity,
where general coordinate transformations are allowed and all observers
at the same space-time point do not always have access to identical regions
of space-time---one observer may find part of the space-time blocked
by a horizon, while another may not see any horizon.
In the broadest sense, quantum general covariance will demand
a democratic treatment of all observers,
irrespective of the limited space-time access he/she may have.
An effective theory for any observer can be constructed by integrating
out the degrees of freedom inaccessible to him/her.
This leads to important constraints on the nature of the gravitational dynamics.
First, the integration must add only total divergence boundary terms
to the effective action, so that the classical Einstein equations of
general relativity remain unaffected.
Second, since this total divergence structure of $A_{\rm boundary}$
is not sufficient to leave the quantum theory of gravity unaffected,
particularly because quantum correlations can exist across a horizon,
we must impose the condition $\exp(iA_{\rm boundary})=1$ to ensure
that the effective theory is protected against quantum fluctuations.

Effective theories are most predictive when they contain only a small number
of terms, because the coefficients of the terms are empirical parameters.
For this purpose, possible terms in the effective theory action are
restricted using symmetry principles and truncated to low orders in the
derivative expansion.
The possible terms that may appear in the effective action for gravity
can be obtained from general principles, as described in Appendix B.
This general analysis shows that only boundary terms appear in the
effective theory, whose parameters have to be empirically determined.
When the boundary is a null surface such as the horizon,
only a unique boundary term survives in the effective action;
in a sense, the principle of effective theory leads to holographic
behaviour for one-way membranes.
It should be noted that the effective theory description will be valid
for any extension of general relativity (supergravity, string theory,
loop quantum gravity, anything else).
Some of the extensions may allow topological changes of space-time,
and the effective theory analysis is fully capable of tackling them.

We have used the ADM formulation of gravity to study the consequences
of the boundary term on the dynamical evolution of space-time.
With a 3+1 foliation covering the whole space-time, the boundary term
can be explicitly obtained by integrating out the inaccessible degrees
of freedom beyond the horizon in the semiclassical approximation.
The quantization condition (\ref{condition}) fixes the normalization
of the boundary term, and produces a uniformly spaced spectrum for it.

It is worth observing that even though the total divergence form of
$A_{\rm boundary}$ and its quantization (\ref{spectrum}) would hold in
the complete quantum theory of gravity, the interpretation of
$A_{\rm boundary}$ in terms of the horizon area holds only in the lowest
order effective theory, and in the semiclassical limit.
Higher order corrections can change the form of $A_{\rm boundary}$
so that it no longer is proportional to the horizon area,
while the true quantum area operator can differ from the
$A_{\rm boundary}$ term which only measures the projection of
the area operator on the horizon surface.
Staying within the lowest order effective theory means that one should
not go very close to the horizon, and semiclassical limit means that
the horizon area parameter $m$ should be large enough.
With the usual power counting counting arguments, these conditions can
be quantified to mean that our result for $A_{\rm boundary}$ is valid
up to ${\cal O}(L^{-1}_{\rm Planck})$ and ${\cal O}(\ln m)$ corrections.

Since the boundary term arises from integrating out the inaccessible
degrees of freedom, it is natural to connect it to the entropy of
the region blocked by the horizon.
Analytical continuation of the boundary term to Euclidean time
confirms this expectation, and various thermodynamic relations
describing properties of the horizon follow.
In our framework, the horizon entropy arises purely from integrating out
the gravitational degrees of freedom, and it is highly tempting to interpret
the discrete value of the boundary term as the result of quantum topological
changes of the region hidden by the horizon.
To really discover the quantum topological features of gravity,
we need to go beyond the framework presented here,
and that is under investigation \cite{inprogress}.

\section*{Acknowledgments}

One of us (TP) thanks K. Subramanian and Tulsi Dass for comments
on the earlier draft of the paper.

\appendix
\section{The Structure of the Action Functional}

We foliate the space-time by a series of space-like hypersurfaces $\Sigma$
with normals $u^i$. From the relation
$R_{abcd}u^d = (\nabla_a \nabla_b - \nabla_b \nabla_a) u_c$,
we obtain
\begin{eqnarray}
R_{bd} u^b u^d &=& g^{ac} R_{abcd} u^b u^d
 = u^b \nabla_a \nabla_b u^a - u^b \nabla_b \nabla_a u^a \nonumber \\
&=& \nabla_a(u^b \nabla_b u^a) - (\nabla_au^b)(\nabla_b u^a)
   -\nabla_b(u^b \nabla_a u^a) + (\nabla_b u^b)^2 \nonumber \\
&=& \nabla_i(Ku^i +a^i) - K_{ab}K^{ab} + K_a^a K^b_b
\label{rabcd}
\end{eqnarray}
(Note that for $K_{ij} = K_{ji} = -\nabla_i u_j - u_i a_j$,
we have $K \equiv K^i_i = -\nabla_i u^i$ and
$K_{ij} K^{ij} = (\nabla_iu^j) (\nabla_ju^i)$).
Further using
\begin{equation}
R = -R~g_{ab} u^a u^b = 2 (G_{ab}-R_{ab}) u^a u^b ~,
\label{rinhone}
\end{equation}
and the identity
\begin{equation}
2~G_{ab} u^a u^b = K_a^a K^b_b   - K_{ab}K^{ab} + {}^{(3)}{\cal R}~,
\label{ginr}
\end{equation}
we can write the scalar curvature as
\begin{equation}
R = {}^{(3)}{\cal R} +K_{ab}K^{ab} - K_a^a K^b_b - 2 \nabla_i (Ku^i + a^i)
  \equiv L_{\rm ADM} -2 \nabla_i (Ku^i + a^i) ~,
\label{rinh}
\end{equation}
where $L_{\rm ADM}$ is the ADM Lagrangian.
This is the result used in the article.
   
Let us now integrate (\ref{rinh}) over a four volume ${\cal V}$
bounded by two space-like hypersurfaces $\Sigma_1$ and $\Sigma_2$
and a time-like hypersurface ${\cal S}$.
The space-like hypersurfaces are constant time slices with normals $u^i$,
and the time-like hypersurface has normal $n^i$ orthogonal to $u^i$.
The induced metric on the space-like hypersurface $\Sigma$ is
$h_{ab} = g_{ab} + u_a u_b$, while the induced metric on the time-like
hypersurface ${\cal S}$ is $\gamma_{ab} = g_{ab} - n_a n_b$.
$\Sigma$ and ${\cal S}$ intersect along a 2-dimensional surface ${\cal Q}$,
with the induced metric
$\sigma_{ab} = h_{ab} - n_a n_b = g_{ab} + u_a u_b - n_a n_b$. 
With $g_{00}=-N^2$, we get
\begin{eqnarray}
A_{\rm EH} &=& {1 \over 16\pi} \int_{\cal V} d^4x~\sqrt{-g}~R \nonumber \\
  &=& {1 \over 16\pi} \int_{\cal V} d^4x~\sqrt{-g}~L_{\rm ADM}
    -{1 \over 8\pi} \int_{\Sigma_1}^{\Sigma_2} d^3x~\sqrt{h}~K
    - {1 \over 8\pi} \int_{\cal S}
      dt~d^2x~N~\sqrt{\sigma}(n_ia^i) ~.
\label{bigeqn}
\end{eqnarray}
 
Let the hypersurfaces $\Sigma, {\cal S}$ as well as their intersection
2-surface ${\cal Q}$ have the corresponding extrinsic curvatures
$K_{ab}, \Theta_{ab}$ and $q_{ab}$. 
In the literature, the Einstein-Hilbert action is conventionally expressed
as a term having only the first derivatives, plus an integral of the trace
of the extrinsic curvature over the bounding surfaces.
It is easy to obtain this form using the foliation condition $n_i u^i=0$
between the surfaces, and noting
\begin{equation}
n_i a^i = n_i u^j \nabla_j u^i = -u^j u^i \nabla_j n_i
= (g^{ij} - h^{ij}) \nabla_jn_i =- \Theta + q ~,
\label{thetaink}
\end{equation} 
where $\Theta\equiv\Theta^a_a$ and $q\equiv q^a_a$ are the traces of the
extrinsic curvature of the surfaces, when treated as embedded in the
4-dimensional or 3-dimensional enveloping manifolds.
Using (\ref{thetaink}) to replace $(n_i a^i)$ in the last term of
(\ref{bigeqn}), we get the result   
\begin{eqnarray}
A_{\rm EH} &+& {1 \over 8\pi} \int_{\Sigma_1}^{\Sigma_2} d^3x~\sqrt{h}~K
 - {1 \over 8\pi} \int_{\cal S} dt~d^2x~N~\sqrt{\sigma}~\Theta \nonumber \\
&=& {1 \over 16\pi} \int_{\cal V} d^4x~\sqrt{-g}~L_{\rm ADM}
 - {1 \over 8\pi} \int_{\cal S} dt~d^2x~N~\sqrt{\sigma}~q ~.
\label{EHtoADM}
\end{eqnarray}
In the first term on the right hand side,
$L_{\rm ADM}$ contains ${}^{(3)}{\cal R}$,
which in turn contains second derivatives of the metric tensor.
The second term on the right hand side removes these second derivatives
making the right hand side equal to the $\Gamma^2$-action for gravity.
On the left hand side, the second and third terms are integrals of the
extrinsic curvatures over the boundary surfaces, which when added to the
Einstein-Hilbert action give the quadratic action without second derivatives.
This is the standard result often used in the literature,
which---unfortunately---misses the importance of the $(n_i a^i)$ term
in the action by splitting it.

\section{Possible Boundary Terms in the Effective Action for Gravity}

For any field theory, its symmetry principles constrain the structure
of terms that may appear in its action.
In case of gravity, the symmetry principle is general covariance.
When the space-time has no boundary, allowed terms in the Lagrangian
density for gravity are invariant scalars formed from the metric
$g_{ij}$ and the covariant derivative operator $\nabla_k$.
The action is obtained by integrating these invariant scalars over the
the invariant space-time measure $d^4x \sqrt{-g}$.
When the effective action is restricted to contain no more than two
derivative operators, there are only two possible terms,
a constant and $R$, and the action takes the form
\begin{equation}
A_{\rm bulk} \equiv  \int d^4x~\sqrt{-g}~(c_1+c_2R) ~=
{1\over 16\pi G} \int d^4x~\sqrt{-g}~(R-2\Lambda) ~,
\end{equation}  
which is the Einstein-Hilbert action with a cosmological constant.
The constants $c_1$ and $c_2$ are traded of for $G$ and $\Lambda$,
but their values---not even their signs---cannot be ascertained
without further physical inputs. 

When the space-time has a boundary,
additional variables describing the geometry of the boundary
can also appear in the action, giving rise to new terms.
The form of these new terms can still be restricted to a great
extent by general considerations as described below.

\subsection{Euclidean gravity}

The structure of the boundary terms is easier to understand
in Euclidean space-time. Let $\mathcal{M}$ be a compact region
of Euclidean space-time with the boundary $\partial\mathcal{M}$.
The geometry of a closed orientable boundary can be fully described
by its outward pointing unit normal, $w^a$, with $w^aw_a=1$.
Generically the boundary is specified in a coordinate dependent way
(e.g. by setting some coordinate to a constant value).
Then general coordinate transformations move the boundary around,
and $w^a$ is not a generally covariant vector,
although it is Lorentz covariant (with appropriate Euclidean meaning).
In such a situation, possible terms in the effective action for gravity
can be obtained by fictitiously treating $w^a$ as a generally covariant
vector, and constructing all possible invariant scalars.
This prescription ensures that symmetries of the effective action
are violated only through $w^a$, and not through any other variable.
(For example, a similar prescription is used to construct effective
chiral Lagrangians of strong interactions. The chirally non-invariant
mass parameter is assigned a fictitious transformation property that
cancels the chiral transformation of hadron fields,
and then all possible chirally invariant terms are written down.)

Since $w^a$ is defined only on the boundary,
the action terms containing it have to be boundary integrals.
Since we would like to have a local Lagrangian description of the theory,
a foliation may be used to extend the value of $w^a$ through the whole
space-time. Such a foliation is by no means unique, and the action
terms must not depend on how the extension of $w^a$ is carried out.
This requirement is guaranteed by demanding that action terms
containing $w^a$ be total divergences; Gauss's law then implies that
$w^a$ defined on the boundary is sufficient to evaluate these terms.
Thus the boundary terms in the effective action for gravity take the form
\begin{equation}
A_{\rm boundary} = ({\rm constant}) \int d^4x \sqrt{g}~\nabla_a V^a,
\end{equation}
where $V^a$ is a vector constructed from $g_{ij}$, $\nabla_k$ and $w_b$.
This total divergence term does not affect the dynamics of classical
gravity, but it may affect the dynamics of the quantum theory.

With the restriction of no more than two derivative operators in the action, 
the possible candidates for $V^a$ are:
(i) with zero derivatives, $V^a=w^a$;
(ii) with one derivative, 
$V^a=(w^a \nabla_b w^b, w^b \nabla_b w^a, w_b \nabla^a w^b)$.
Of these, $w_b \nabla^a w^b = 0$, due to the normalization $w^b w_b=1$.
Furthermore, the term $\nabla_j (w^b \nabla_b w^j)$ involving the
``acceleration" $a^j= w^b \nabla_b w^j$ integrates to zero,
since use of Gauss's law converts the integrand to $w_ja^j$
which vanishes identically.
Thus we are left with just two possibilities $V^a=(w^a,w^a \nabla_b w^b)$.
In terms of the induced metric on $\partial\mathcal{M}$, $f_{ab}$,
the corresponding contributions to the action are:
\begin{equation}
A_{\rm boundary}^{(1)}
= \int_{\mathcal{M}} d^4x \sqrt{g}~\nabla_a w^a
= \int_{\partial\mathcal{M}} d^3x \sqrt{f}
= {\rm Vol}(\partial\mathcal{M}) ~,
\label{volumeterm}
\end{equation}
\begin{equation}
A_{\rm boundary}^{(2)}
= \int_{\mathcal{M}} d^4x \sqrt{g}~\nabla_a (w^a \nabla_b w^b)
= \int_{\partial\mathcal{M}} d^3x \sqrt{f}~\nabla_b w^b ~
= -\int_{\partial\mathcal{M}}d^3x \sqrt{f}~K_{\rm ext} ~.
\end{equation}
The former is the total volume of the boundary,
while the latter is the integral of the extrinsic curvature
$K_{\rm ext}=-\nabla_b w^b$ over the boundary.
The most general lowest order effective action
for Euclidean gravity thus takes the form
\begin{equation}
A_{\rm grav} \equiv \int_{\mathcal{M}} d^4x~\sqrt{g}~(c_1+c_2R)
  + \int_{\partial\mathcal{M}} d^3x \sqrt{f}~(c_3+c_4 K_{\rm ext}) ~.
\label{euclsurface}
\end{equation}  
The constants appearing here have the dimensions
$c_1\sim L^{-4}, c_2\sim L^{-2}, c_3\sim L^{-3}, c_4\sim L^{-2}$.

For a closed but only piecewise smooth boundary,
it is convenient to separate $A_{\rm boundary}$ in to the contribution
from the smooth part and the contribution from the edges.
The normal to the boundary is discontinuous in going across the edge,
and so the gradient of the normal is singular along the edge.
This singularity does not contribute to $A_{\rm boundary}^{(1)}$,
but it does contribute to $A_{\rm boundary}^{(2)}$.
The edge contribution to $A_{\rm boundary}^{(2)}$ can be evaluated
by rounding off the edge in a limiting procedure.
In this limit, the curvature cancels with the corresponding factor from
the integration measure, and only the angular discontinuity $\delta$
of the normal across the edge is left behind.
Let $\mathcal{Q}$ be the set of edges with the induced metric $\sigma_{ab}$.
Then
\begin{equation}
-\int_{\partial\mathcal{M}}d^3x \sqrt{f}~K_{\rm ext} ~\longrightarrow~
-\int_{\partial\mathcal{M},{\rm smooth}}d^3x \sqrt{f}~K_{\rm ext}
+\int_{\mathcal{Q}}d^2x \sqrt{\sigma}~\delta ~.
\end{equation}
A common situation is the one where the unit normals to the boundary
on either side of the edge, $w^{(1)a}$ and $w^{(2)a}$, are orthogonal.
The edge contribution involves both, in the form of a double divergence.
The first divergence embeds the 3-dimensional boundary in the 4-dimensional
space-time and then the second divergence embeds the 2-dimensional edge
in the 3-dimensional boundary.
With $D$ denoting the projection of $\nabla$ on to the boundary
$\partial\mathcal{M}$, the edge contribution becomes
\begin{eqnarray}
A_{\rm boundary}^{(2),{\rm edge}}
= {\pi\over2}
  \int_{\mathcal{M}} d^4x \sqrt{g}~\nabla_a (w^{(1)a} f_{bc} D^b w^{(2)c})
= {\pi\over2} \int_{\partial\mathcal{M}} d^3x \sqrt{f}~f_{bc} D^b w^{(2)c}
= {\pi\over2} \int_{\mathcal{Q}} d^2x \sqrt{\sigma} ~,
\end{eqnarray}
which is the total area of the edge surface multiplied by the angular
discontinuity $\delta=\pi/2$.

If the boundary can be analytically continued back to Minkowski
space-time in some suitable coordinates, then we will have
analogous boundary terms in Minkowski space-time also.
If we impose the condition that the complete action does not contain
derivatives higher than the first, then we must have $c_4=2c_2$.
We see from equation (\ref{EHtoADM}) that in this case the sum of
bulk and surface terms lead to the quadratic ($\Gamma^2$) action.
We stress that the ADM action contains second derivatives of $g_{ab}$
with respect to the spatial coordinates (through ${}^{(3)}R$),
while the $\Gamma^2$ action has no second derivatives at all.
In Euclidean space-time, both space and time are treated on an equal
footing, and any covariant prescription which removes the second
derivatives along one direction will remove them along all directions,
leading to the $\Gamma^2$ action.

\subsection{Minkowski space-time with foliation}

It is possible to figure out the possible boundary terms even for a
space-time with Minkowski signature, in presence of pre-specified foliations.
In this case, the boundaries typically have both space-like ($\Sigma$)
and time-like (${\cal S}$) parts as described in Appendix A.
The corresponding normals $u^a$ and $n^a$ satisfy $u_au^a=-1, n_an^a=1$,
and $u^an_a=0$ on the intersection of $\Sigma$ and ${\cal S}$.
$\Sigma$ and ${\cal S}$ have induced metrics $h_{ab}=g_{ab}+u_a u_b$
and $\gamma_{ab}=g_{ab}-n_a n_b$, and extrinsic curvatures $K=-\nabla_a u^a$
and $\Theta=-\nabla_a n^a$ respectively.
Now we need to make a clear distinction between two different situations: 

(a) We may insist that $u^a$ is given only on $\Sigma$,
and $n^a$ is given only on $\mathcal{S}$ with arbitrary extensions elsewhere.
Then the surface terms we obtain should not depend on the manner in which
these are extended.
This situation is similar to that of the Euclidean case discussed above,
with one crucial difference:
The surface of intersection of $\Sigma$ and ${\cal S}$ (i.e. $\mathcal{Q}$)
cannot be smoothly rounded off because the two normals $u^a$ and $n^a$ have
normalizations of opposite signs.
With unrelated coefficients for the space-like and the time-like parts of
the boundary, the boundary action becomes
\begin{equation}
A_{\rm boundary}
= k_1{\rm Vol}(\Sigma) + k_2\int_\Sigma d^3x \sqrt{h}~K
+ k_3{\rm Vol}(\mathcal{S}) + k_4\int_\mathcal{S}dtd^2x \sqrt{|\gamma|}~\Theta
+ k_5{\rm Area}(\mathcal{Q}) ~.
\label{finalnofoli}
\end{equation}
If a smooth analytic continuation between Minkowski and Euclidean
space-times is assumed, then the coefficients are related according to:
$k_3=-k_1, k_4=-k_2, k_5=(\pi/2)k_2$.

(b) It is, however, commonplace to assume a more elaborate geometry, viz.
that there exists a \emph{foliation} of space-time by space-like surfaces, 
for which $u^a(x)$ is the normal.
For example, the ADM formulation explicitly uses such a foliation.
In such a case, the possible terms in the action can depend on the
complete vector field $u^a(x)$ and not just its value on the boundary.
The resultant action would have more parameters, and hence would be less
predictive.
In general, the space-time structure need not admit a foliation
in terms of time-like surfaces,
and we will continue to assume that $n^a$ is specified \emph{only} on $\mathcal{S}$.
Our final results should then depend only on the value of $n^a$ on the
boundary, and not in the manner in which it may be extended elsewhere.

Let us proceed as in the case of Euclidean space-time,
pretending that $u^a$ and $n^a$ can be extended
in some sensible fashion to the bulk as two vector fields.
With $u^a(x)$ and $n^a(x)$ treated as genuine vector fields,
one can write down several new terms for the bulk action.
We, however, focus on total divergence boundary terms only.
These terms are again integrals of the form $\nabla_a V^a$, where $V^a$ is
built from $g_{ij}, \nabla_k, u^b, n^c$, with the derivative acting at most once.  
The following vectors exhaust the possibilities:
\begin{eqnarray}
V^a = &(& u^a, u^a\nabla_b u^b, u^b\nabla_b u^a,
               u^a\nabla_b n^b, u^b\nabla_b n^a, \nonumber \\
       && n^a, n^a\nabla_b n^b, n^b\nabla_b n^a,
               n^a\nabla_b u^b, n^b\nabla_b u^a, \nonumber \\
       && u^b\nabla^a n_b, u^au^bu^c\nabla_b n_c, u^an^bu^c\nabla_b n_c,
                           n^au^bu^c\nabla_b n_c, n^an^bu^c\nabla_b n_c\ ) ~.
\label{listofv}
\end{eqnarray}
The second line is obtained from the first, by the obvious interchange
$u^a \longleftrightarrow n^a$. 
For the third line, there is no need to include  such interchanges,
 because $u^b\nabla^an_b = -n_b\nabla^au^b$.
The first three terms of both first and second lines have already been
discussed in the previous subsection, and the remaining terms arise
because of the existence of the second vector field.
When $\nabla_a V^a$ is integrated over a space-time region,
we get (i) boundary integral of $u_a V^a$ over $\Sigma$,
and (ii) boundary integral of $n_a V^a$ over $\mathcal{S}$. 

These dot products are given by
\begin{equation}
u_a V^a = (1, \nabla_au^a, 0, \nabla_an^a, u^au^b\nabla_bn_a,\ \ 
           0, 0, u^an^b\nabla_bn_a, 0, 0,\ \ 
           u^au^b\nabla_an_b, u^au^b\nabla_an_b, n^au^b\nabla_an_b, 0, 0) ~,
\label{udotv}         
\end{equation}
and
\begin{equation}
n_a V^a = (0, 0, n^au^b\nabla_bu_a, 0, 0,\ \ 
           -1, -\nabla_an^a, 0, -\nabla_au^a, n^an^b\nabla_bu_a,\ \ 
           n^au^b\nabla_an_b, 0, 0, -u^au^b\nabla_an_b, -n^au^b\nabla_an_b) ~.
\label{ndotv}         
\end{equation}
Together, we have to consider the set of scalars
\begin{eqnarray}
(u_aV^a, n_aV^a) =
(1, \nabla_au^a, \nabla_an^a, u^an^b\nabla_bn_a, n^au^b\nabla_bu_a) ~,
\label{fullset}
\end{eqnarray}
where we have used the condition $n_au^a=0$ to eliminate some of the
scalars in favour of the ones listed.

Let us begin with possible integrals over the boundaries $\Sigma$.
We first note that the vector $n^a$ is \emph{not} specified over most of the
surface (it exists only because we arbitrarily extended it for convenience),
but it is known on the 2-surface $\mathcal{Q}$.
This, in turn, implies that we cannot have the two terms
$(u^an^b\nabla_bn_a,  n^au^b\nabla_bu_a)$ integrated over $\Sigma$,
since their values depend on how $n^a$ is extended.
The $\nabla_a n^a$ term requires more care, however.
Note that the projection $D_an^a$ of $\nabla_an^a$ on to $\Sigma$ is given by
\begin{equation}
D_an^a \equiv (\delta^i_a+u^iu_a) (\delta_j^a+u_ju^a) \nabla_in^j
       = \nabla_an^a +u^iu^j\nabla_in_j ~.
\label{dproject}
\end{equation}
Therefore, $\nabla_an^a$ differs from $h^{ab}D_an_b$ only by the term
$u^au^b\nabla_an_b = -n_bu^a\nabla_au_b$, which has already been considered.
The integral of $h^{ab}D_an_b$ over $\Sigma$ is allowed, because Gauss's law
converts it in to an integral over $\mathcal{Q}$ on which $n^a$ is known.
(In other words, the integral of either of the two terms on the right hand
side of (\ref{dproject}) depends on the manner in which $n_a(x)$ is extended
in to $\Sigma$, but their combined integral depends only on the known value
of $n_i$ on $\mathcal{Q}$.)
The corresponding boundary action is
\begin{equation}
A_{\rm boundary}^{(3)}
= -\int_\Sigma d^3x \sqrt{h}~h_{bc} D^b n^c
= -\int_{\mathcal{Q}} d^2x \sqrt{\sigma},
\end{equation}
which is the total area of the surface $\mathcal{Q}$.
Thus the total contribution of terms in (\ref{fullset}) on the boundary
$\Sigma$ is
\begin{equation}
A_{\rm boundary:\Sigma}
= \int_\Sigma d^3x \sqrt{h}~(k_1+k_2 K) 
  + k_5\int_{\mathcal{Q}} d^2x \sqrt{\sigma}
= k_1{\rm Vol}(\Sigma) + k_2\int_\Sigma d^3x \sqrt{h}~K
  + k_5{\rm Area}(\mathcal{Q})
\end{equation}
If the total action has to contain only up to first {\it time} derivatives,
then $k_2=2c_2$. 

Let us next consider the contribution of the terms in (\ref{fullset})
on the boundary $\mathcal{S}$.
The first three terms can be analyzed just as in case of $\Sigma$, yielding
\begin{equation}
A_{\rm boundary:\mathcal{S}}
= \int_{\mathcal{S}}  d^3x \sqrt{|\gamma|}~(k_3+k_4 \Theta) 
  + k_5\int_{\mathcal{Q}} d^2x \sqrt{\sigma}
= k_3{\rm Vol}(\mathcal{S}) + k_4\int_{\mathcal{S}}  d^3x \sqrt{|\gamma|}~\Theta
  + k_5{\rm Area}(\mathcal{Q})
\end{equation}
In addition, since $u^a(x)$ is a vector field defined everywhere in
$\mathcal{S}$, we can no longer ignore the last two terms
$(u^an^b\nabla_bn_a, n^au^b\nabla_bu_a)$ while integrating over $\mathcal{S}$.
(These are the ``accelerations" of one normal dotted with the other normal.)
Of these, the term $n^au^b\nabla_bu_a=n^ia_i$ leads to the integral of the
surface gravity, and has been extensively discussed in earlier sections of the paper:
\begin{equation}
A_{\rm boundary}^{(4)}
= k_6\int_\mathcal{S}dtd^2x \sqrt{|\gamma|}~n^ia_i
= k_6\int_\mathcal{S}dtd^2x N\sqrt{\sigma}~n^ia_i ~.
\label{ndota}
\end{equation}
What remains is the term
\begin{equation}
A_{\rm boundary}^{(5)}
= k_7\int_\mathcal{S}dtd^2x \sqrt{|\gamma|}~u^an^b\nabla_bn_a
=-k_7\int_\mathcal{S}dtd^2x N\sqrt{\sigma}~n^an^b\nabla_bu_a ~.
\label{mystery}
\end{equation}
To understand the nature of this term, consider a coordinate system
in which $x^1=$ constant corresponds to the surface $\mathcal{S}$,
and the metric has the form $g_{00}=-N^2, g_{11}=M^2, g_{0\alpha}=0,
g_{1A}=0, g_{AB}=\sigma_{AB}$ with $A,B=2,3$.
Then, a simple calculation shows that
$n^an^b\nabla_bu_a=(MN)^{-1}(\partial M/\partial t)$.
This term vanishes if $(\partial M/\partial t)=0$.
In general, its contribution is
\begin{equation}
A_{\rm boundary}^{(5)}
= -k_7\int_\mathcal{S}dtd^2x N\sqrt{\sigma}~\frac{\dot M}{MN}
= -k_7\int_\mathcal{Q}d^2x \sqrt{\sigma}~(\ln M) \bigg|_{t_1}^{t_2} ~.
\end{equation}

To summarize, equation (\ref{finalnofoli}) gives the total boundary action,
when $u^i$ and $n^i$ are defined only on the boundary and no foliation is
assumed.
This result agrees with the one obtained in case of Euclidean space-time.
If the vector field $u^a(x)$ is related to a space-time foliation,
then two additional terms appear, given by (\ref{ndota}) and (\ref{mystery}).
This concludes our general analysis of boundary terms.

\subsection{Boundary terms for a horizon}

When the boundary is a horizon, and not just any hypersurface,
the possible terms in the effective action get restricted further.
This happens because the integration measure in time direction is
$Ndt$, and with $N=0$ on the horizon some of the integrals vanish.
Among the various contributions in (\ref{finalnofoli}),
(i) The term ${\rm Vol}(\mathcal{S})$ vanishes.
(ii) In the term with $\Theta$, we can use (\ref{thetaink}) to write
$\Theta=q-n^ia_i$, and note that the integral of $q$ vanishes.
The $n^ia_i$ term, which of course is the same as in (\ref{ndota}),
leads to the area of the horizon when $N\to0$.
(iii) Finally, the term in (\ref{mystery}) vanishes when $N\to0, MN=$constant,
or when  $\dot{M}=0$  on the horizon.
Thus, when the boundary is a horizon, the most general boundary action
is of the form
\begin{equation}
A_{\rm boundary: horizon}
= k_1{\rm Vol}(\Sigma) + k_2\int_\Sigma d^3x \sqrt{h}~K
+ k_4{\rm Area}(\mathcal{Q}) ~.
\end{equation}
The result obtained with the ADM action in the paper corresponds to
$k_1=0, k_2=2c_2,k_4=4\pi c_2$.


\begin{thebibliography}{99}
\bibitem{comment}
Integration over null surfaces like horizons requires some care,
and can be handled using a limiting procedure.
One also needs to be careful in handling the intersection of future and past
space-like surfaces $\Sigma_2$ and $\Sigma_1$ on the bifurcation sphere.
Though we have not discussed these issues in detail,
they can be handled rigorously and do not affect our arguments or results.
\end{thebibliography}
\end{document}